\documentclass[a4paper,11pt]{article}

\usepackage[utf8]{inputenc}
\usepackage[T1,T2A]{fontenc}
\usepackage{amsmath,amsfonts,amssymb}

\usepackage{graphicx}
\usepackage{cite}
\usepackage{hyperref}

\usepackage{geometry}
\allowdisplaybreaks

\begin{document}

\begin{titlepage}

\begin{center} \textbf{\LARGE Dark matter production via a non-minimal coupling to gravity} \end{center}

\vspace{1cm}

\begin{center}
  \textbf{Oleg Lebedev\(^a\), Timofey Solomko\(^b\), Jong-Hyun Yoon\(^c\)}
\end{center}

\begin{center}
  \vspace*{0.15cm}
  \textit{\({}^a\)Department of Physics and Helsinki Institute of Physics,\\
  Gustaf H\"allstr\"omin katu 2a, FI-00014 Helsinki, Finland}\\
  \vspace*{0.15cm}
  \textit{\({}^b\)Saint Petersburg State University, 7/9 Universitetskaya nab.,\\
  St.Petersburg, 199034, Russia}\\
  \vspace*{0.15cm}
  \textit{\({}^c\)Université Paris-Saclay, CNRS/IN2P3, IJCLab, 91405 Orsay, France}
\end{center}

\vspace{2.5cm}

\begin{center} \textbf{Abstract} \end{center}

\noindent We study postinflationary scalar dark matter production via its non-minimal
coupling to gravity. During the inflaton oscillation epoch, dark matter is produced
resonantly for a sufficiently large non-minimal coupling \(\xi\gtrsim 5\). We find that
backreaction on the curvature and rescattering effects typically become important for the
values of \(\xi\) above \(30\), which invalidate simple estimates of the production
efficiency. At large couplings, the dark matter yield becomes almost independent of
\(\xi\), signifying approximate quasi-equilibrium in the inflaton-dark matter system.\
Although the analysis gets complicated by the presence of apparent negative energy in the
Jordan frame, this behaviour can be regularized by introducing mild dark matter
self-interaction.\ Using lattice simulations, we delineate parameter space leading to the
correct dark matter relic abundance.

\end{titlepage}

\tableofcontents

\section{Introduction}

Inflation is a cornerstone of modern cosmology, which explains many observed features of
the Universe~\cite{Starobinsky:1980te,Guth:1980zm,Linde:1981mu,Linde:1983gd,Mukhanov:1981xt}.\
It also entails significant particle production of various kinds. In particular,
inflationary and postinflationary particle production is an important source of stable
relics, which may contribute to dark matter. It occurs due to non-adiabatic variation
of the background field, e.g. the metric~\cite{Parker:1969au,Grib:1969ruc} or the
inflaton~\cite{Dolgov:1989us,Traschen:1990sw} field. Recent reviews and discussions of
the subject can be found in~\cite{Ford:2021syk,Lebedev:2021xey}.

In this work, we study postinflationary production of real scalar dark matter $s$ via
its non-minimal coupling to curvature~\cite{Chernikov:1968zm,Buchbinder:1992rb},
\begin{equation}
  \Delta \mathcal{L}_\xi = -{1\over 2} \xi R \, s^2 \;,
\end{equation}
where $\xi$ is a real constant. As $R$ starts oscillating after inflation, the effective
mass term for $s$ changes its sign leading to efficient particle production. We assume
that $s$ is stable and very weakly coupled such that it constitutes \textit{non-thermal}
dark matter (DM).\footnote{One may entertain the possibility that the inflaton itself
plays the role of dark matter. However, this option is not available for non-thermal
dark matter, at least in the minimal case~\cite{Lebedev:2021zdh}.}

Particle production via a non-minimal coupling to gravity has been studied in a number of
papers starting with~\cite{Bassett:1997az}. Most work was focused on the linear regime,
possibly including the Hartree approximation~\cite{Tsujikawa:1999iv,Bertolami:2010ke,Koivunen:2022mem}.
Dark matter with a non-minimal coupling to gravity was considered
in~\cite{Markkanen:2015xuw,Fairbairn:2018bsw,Cembranos:2019qlm,Babichev:2020yeo,Clery:2022wib} within the
same approximation.\footnote{See also~\cite{Cata:2016dsg} for effects of the non-minimal
DM coupling in a different context.} Recently, this treatment has been improved
in~\cite{Kainulainen:2022lzp} by employing the quantum transport equations.

In our work, we make use of lattice simulations to incorporate important collective
effects in dark matter production. We find that these crucially affect the resulting
abundance, especially at large $\xi$. Such effects include significant backreaction of
the produced particles on the curvature as well as rescattering, which lead to
quasi-equilibrium in the inflaton-DM system at large $\xi$. In this regime, the dark
matter abundance becomes almost independent of $\xi$. Analogous behaviour has been
observed for direct inflaton-DM coupling in~\cite{Lebedev:2021tas,Lebedev:2022ljz}.

We perform our computations in \textit{locally} quadratic and quartic inflaton potentials,
focussing on positive $\xi $ such that the field $s$ is sufficiently heavy during
inflation and not subject to large de Sitter fluctuations. This allows us to use the
vacuum state of the dark matter field as the initial condition for our simulations. In
contrast, a negative $\xi$ leads to a large VEV of $s$ during inflation, which affects
its eventual abundance. For comparison, however, we also display our results for $\xi<0$
with vacuum initial conditions, whenever possible. Finally, we delineate parameter space
leading to the correct relic abundance of non-thermal dark matter.

\section{Dynamics in the Jordan frame}\label{jordan_dynamics}

In the Friedmann metric
\begin{equation}
ds^2 = -dt^2 + a(t)^2 \, dx^i \,dx^i \;,
\end{equation}
the scalar curvature is given by
\begin{equation}
R = 6 \left[    {\ddot a \over a} + \left( {\dot a \over a}\right)^2    \right]\;,
\label{R}
\end{equation}
where the dot indicates the time derivative.
We take the action to be of the form
\begin{equation}
\mathcal{S}= \int d^4 x \sqrt{-g} \left( {1\over 2} M_\textrm{Pl}^2 R -  {1\over 2}\xi  R\, s^2 -{1\over 2} g^{\mu\nu} \partial_\mu s \, \partial_\nu s - {1\over 2} g^{\mu\nu} \partial_\mu \phi \, \partial_\nu \phi -V
\right)\;,
\label{action}
\end{equation}
where $s$ is the dark matter field, $\phi$ is the inflaton and $g$ is the metric
determinant. We assume that the potential contains no direct coupling between $s$ and
$\phi$, and can be approximated \textit{after inflation} by
\begin{align}
V= V(\phi) + V(s) &= {1\over 2} m_\phi^2  \phi^2+ {1\over 4} \lambda_\phi \phi^4\nonumber\\
&+ {1\over 2} m_s^2 s^2 + {1\over 4} \lambda_s s^4 \;,
\end{align}
with $m_\phi \gg m_s$. We consider both possibilities that the inflaton potential during
preheating is dominated by the quadratic or quartic parts.

In our study, we perform calculations in the Jordan frame following~\cite{Figueroa:2021iwm},
that is, keeping explicit the coupling $s^2 R$. This leads to simpler equations of motion
compared to those in the Einstein frame. Indeed, the scalar kinetic terms in the Einstein
frame are field-dependent and thus contain fast oscillating functions. This makes the
numerical integration of the equations of motion relatively unstable compared to that in
the Jordan frame. At late times, when the characteristic field value of $s$ becomes small,
the Jordan and Einstein frames become essentially indistinguishable. As a result, the
relevant observables such as (conserved) particle numbers can be read off directly from
the simulation output in the Jordan frame. In practice, we terminate our simulation at
$a/a_0 \sim 10^3$ and compute the observables at that time.

An important aspect of the dynamics in the Jordan frame is the scalar mixing with the
gravitational degrees of freedom~\cite{Salopek:1988qh}. In particular, the scale factor
corresponds to a scalar component of the metric $g^{\mu\nu}$. Using~\eqref{R} and
integrating by parts, one finds the following kinetic terms:
\begin{equation}
\mathcal{K}= {a^3 \over 2} \, \dot s^2 + 6\xi a^2 s \,\dot a \dot s + 3a (\xi s^2 - M_\textrm{Pl}^2)\,\dot a^2 \;.
\label{kin-mix}
\end{equation}
We observe that there is kinetic mixing between $s$ and $a$, which can make a significant
impact. For example, if we take $s$ to scale as radiation $s \sim a^{-1}$ at late times,
the mixing term becomes much larger than the diagonal kinetic term $a^3 \dot s^2$ at
$\xi \gg 1$. On the other hand, the mixing is negligible if $s$ oscillates with a
relatively large frequency $\omega \gg \xi H$ and decreasing amplitude
$\propto a^{-\alpha}$ with $\alpha \sim \mathcal{O}(1)$. In this case, the scalar field
decouples from the metric and one can define a meaningful oscillator number for its
momentum modes.

The Einstein equation is obtained by varying the action with respect to $g^{\mu\nu}$,
\begin{equation}
R_{\mu\nu}- {1\over 2} g_{\mu\nu} R = {1\over M_\textrm{Pl}^2} \, T_{\mu\nu} \;,
\end{equation}
where the energy momentum tensor is
\begin{equation}
T_{\mu\nu} = - {2\over \sqrt{-g}}\, {\delta (\sqrt{-g} \mathcal{L}_\textrm{matter}) \over \delta g^{\mu\nu}} \;.
\end{equation}
The matter Lagrangian $\mathcal{L}_\textrm{matter}$ contains all the terms in the
integrand of~\eqref{action} except for ${1\over 2} M_\textrm{Pl}^2 R$. In particular, the
non--minimal gravity coupling to $s$ is considered part of $\mathcal{L}_\textrm{matter}$,
in which case the energy momentum tensor is covariantly conserved by virtue of the
Einstein equation. Furthermore, since there is no direct coupling between $\phi$ and $s$,
the energy momentum tensor splits naturally into $T_{\mu\nu} (\phi)$ and $T_{\mu\nu}(s)$.
An explicit calculation shows
\begin{equation}
  T_{\mu\nu} (s) = \partial_\mu s \partial_\nu s - g_{\mu\nu} \left(   {1\over 2} g^{\rho \sigma} \partial_\rho s \,\partial_\sigma s + V(s) \right)
  +\xi \left(  R_{\mu\nu}- {1\over 2} g_{\mu\nu} R + g_{\mu\nu} \Box  -\nabla_\mu \nabla_\nu   \right) s^2 \;.
\end{equation}
Here $\Box = g^{\mu\nu } \nabla_\mu \nabla_\nu$ and $\nabla_\mu$ is the covariant
derivative associated with metric $g^{\mu\nu}$. A similar expression holds for
$T_{\mu\nu} (\phi)$ except for the $\xi$--induced term.

The equation of motion for $s$ is obtained by varying the action with respect to $s$,
\begin{equation}
\Box s  -\xi R s - {\partial V \over \partial s} =0 \;,
\label{box-s}
\end{equation}
or
\begin{equation}
\ddot s + 3 {\dot a \over a }\, \dot s - {1\over a^2 } \nabla^2  s + \xi R s + {\partial V \over \partial s}=0\;.
\label{eom-s}
\end{equation}
An analogous equation at $\xi=0$ applies to $\phi$. We note that the term $\xi R$ plays
the role of the induced mass squared for dark matter. After inflation, the scalar
curvature oscillates such that the effective mass of $s$ can turn tachyonic, indicating
particle production.

Since gravity is represented by a single degree of freedom $a(t)$, it is sufficient to
use the trace of the Einstein equation,
\begin{equation}
  R = -{1\over M_\textrm{Pl}^2}  T_{\mu}^\mu \;,
\end{equation}
to solve for the evolution of the system. It simplifies further when one takes a spatial
average of both sides of the equation. Since $T_{\mu\nu}$ itself depends on $R$, solving
for $R$ yields
\begin{equation}
R = {F(s) \over M_\textrm{Pl}^2 } \left[     (1-6\xi) \langle \partial^\mu s \partial_\mu s \rangle +4 \langle V    \rangle
-6 \xi \langle s V^\prime_s \rangle  + \langle  \partial^\mu \phi \, \partial_\mu \phi   \rangle    \right] \;,
\label{R-EOM}
\end{equation}
where
\begin{equation}
F(s)\equiv\frac{1}{1+(6\xi-1)\xi\langle s^2\rangle/M_\text{Pl}^2}  \,.
\end{equation}
Expressing $R$ in terms of $a$ via~\eqref{R}, one obtains an equation of motion for
$a(t)$, which, together with those for $s$ and $\phi$, can be solved numerically.

The initial evolution of $R$ is fully determined by the inflaton field. At later times,
it can be significantly affected by the produced dark matter via the $\xi$-enhanced terms.
In fact, such contributions are instrumental in determining the end of resonant particle
production.

\subsection{Negative energy and kinetic mixing}

Having solved the EOM, one can determine the energy density of dark matter,
\begin{equation}
  \rho (s)= \langle  T_{00} (s) \rangle=\frac{1}{2}\langle\dot{s}^2\rangle+
  \frac{1}{2a^2}\langle(\nabla s)^2\rangle+\langle V(s)\rangle+
  3\xi H^2\langle s^2\rangle+6\xi H\langle s\dot{s}\rangle \;,
  \label{rho(s)}
\end{equation}
where we have dropped the term $\frac{\xi}{a^2}\langle\nabla^2s^2\rangle$, which averages
to zero since $\nabla^2 s^2$ is a total spatial derivative.

The interpretation of $\rho(s)$ is, however, not straightforward due to the presence of
the last two terms. In particular, the term $6\xi H\langle s\dot{s}\rangle$ can be large
and negative at large $\xi$ such that the total energy density becomes negative. For low
frequency modes, this can persist even at asymptotically large times. Indeed, when the
expansion is radiation--like, $R=0$, the late time low--$k$ solution to the EOM with
$V(s)=0$ is $s\propto a^{-1} $. Then, using $\dot s = \dot a \, s^\prime_a$, one finds
negative energy density for $\xi \gg 1$. The origin for this behaviour lies in the
kinetic mixing between the scalar and the metric~\eqref{kin-mix}, which is responsible
for the term $6\xi H\langle s\dot{s}\rangle$. The scalar energy density cannot in general
be separated from that of gravity, however the scalar does ``decouple'' from the metric
at late times when it has mass or self--interaction. This is because such terms decrease
slower in time than the $\xi$-induced contributions do, thereby regularizing negative
$\rho_s$. Since we are interested in small $m_s$, we make use of mild self--coupling
$\lambda_s$, which allows us to define positive, conserved oscillator numbers at the end
of the simulation.

In what follows, we study the system evolution and particle production in the $\phi^4$
and $\phi^2$ inflaton potentials.

\section{Quartic local inflaton potential}

Suppose that the bare inflaton mass $m_\phi$ is small compared to the relevant scales at
the preheating epoch. Then, locally we can approximate the inflaton potential by the
quartic term
\begin{equation}
V(\phi)\simeq {1\over 4 } \lambda_\phi \, \phi^4\;.
\end{equation}
Note that, at late times, when the characteristic momenta redshift to values comparable
to $m_\phi$, the bare inflaton mass starts playing an important role.

In what follows, we study the main features of particle production in this system.

\subsection{The resonance}

Let us start with the case $\lambda_s=0$. The scalar curvature starts oscillating after
inflation, leading to $s$-particle production (see Eq.~\eqref{eom-s}). This process can
be described semi-classically in terms of the tachyonic resonance~\cite{Felder:2000hj}.

Initially, the energy density is dominated by the inflaton zero mode such that the
curvature is given by
\begin{equation}
R = - {T_\mu^\mu (\phi) \over M_\textrm{Pl}^2} = {1\over M_\textrm{Pl}^2 } \left[    4 V(\phi) - \dot \phi^2   \right] \;,
\label{R-V}
\end{equation}
In the quartic potential, the inflaton field is given by~\cite{Greene:1997fu}
\begin{equation}
\phi (t) = \phi_0\, \textrm{cn} \left(  \sqrt{\lambda_\phi} \phi_0 \,t, {1 \over \sqrt{2}}      \right) \;,
\end{equation}
with a slowly varying amplitude
\begin{equation}
\phi_0 \propto 1/a \;.
\end{equation}
According to the convention of~\cite{Greene:1997fu}, cn$(x, 1/\sqrt{2})$ satisfies
$f^{\prime 2} (x)= {1\over 2} \left[1-f^4(x)\right]$. Therefore,
\begin{equation}
R = { \lambda_\phi \phi_0^4 \over 2 M_\textrm{Pl}^2} \, (3 \,\textrm{cn}^4 (\omega t) -1)\;,
\end{equation}
where $\omega=\sqrt{\lambda_\phi}\phi_0$. In this expression, we keep only the leading
term and neglect the derivatives of the scale factor.

Particle production is best analyzed in terms of the spacial Fourier modes of the DM
field $s_k$~\cite{Greene:1997fu}. The equation of motion (Eq.~\eqref{eom-s}) for the
rescaled $k$--modes $X_k = a^{3/2} s_k$ reads
\begin{equation}
\ddot X_k + \left(   {k^2\over a^2} + {9\over 4} w H^2 + \xi R \right) \, X_k=0 \;,
\label{Xk}
\end{equation}
where $V(s)$ is assumed to be negligible and $w$ is the coefficient of the equation of
state, $w=-(1+2 \dot H /(3H^2))$. At large $\xi$, the $wH^2 $ term can be omitted.
Neglecting also the slow time variation of $\phi_0$, we get
\begin{equation}
 X_k^{\prime\prime} +  \left[ \kappa + q \, \textrm{cn}^4(z)\right] \,X_k=0 \;,
\label{ch-4}
\end{equation}
where
\begin{equation}
\kappa = {k^2 \over {\lambda_\phi } \phi_0^2 a^2} - {\xi \phi_0^2 \over 2 M_\textrm{Pl}^2} ~~,~~ q = {3\xi \phi_0^2 \over 2 M_\textrm{Pl}^2} \;,
\end{equation}
$z= \sqrt{\lambda_\phi} \phi_0 \,t$ and the prime denotes differentiation with respect
to $z$. As a result, we have broad tachyonic resonance: $q\gg1$ and $\kappa < 0$ for low
momenta. Note that for a zero mode, $q/\kappa =-3$.

\begin{figure}[t!]
  \begin{center}
    \includegraphics[width=0.49\textwidth]{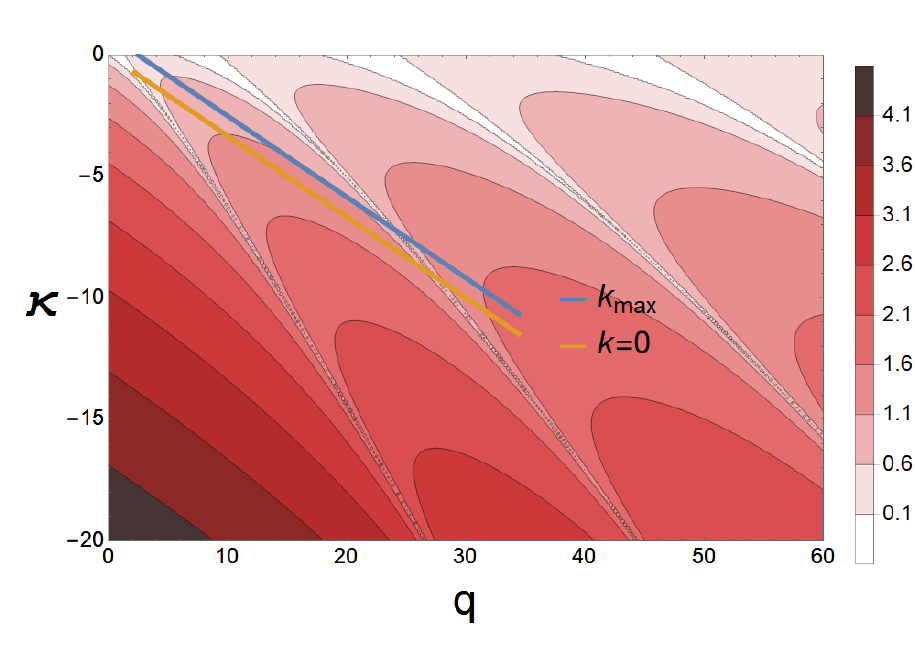}
    \includegraphics[width=0.49\textwidth]{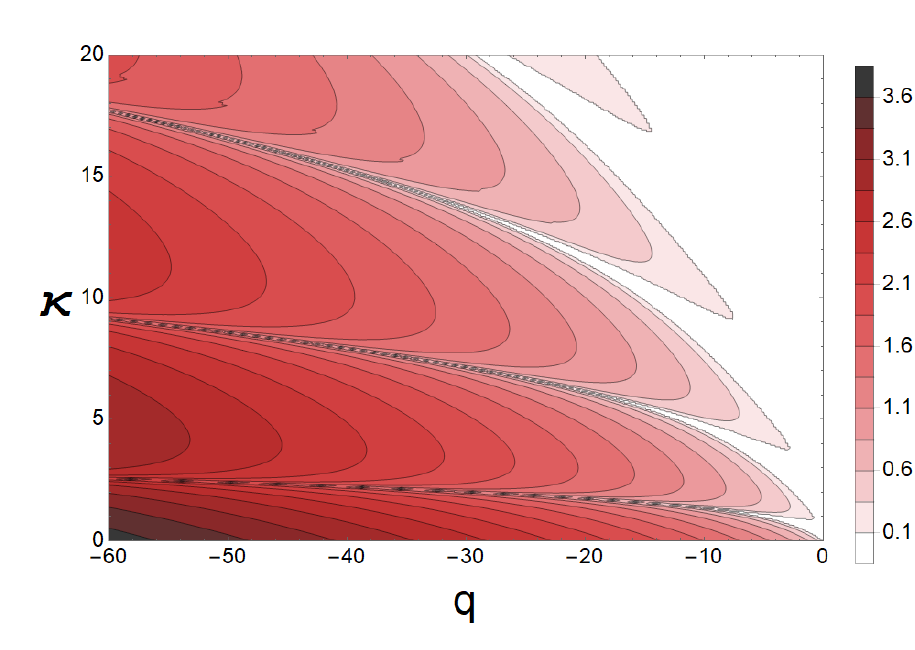}
  \end{center}
  \caption{Stability charts for the ellipsoidal wave equation~\eqref{ch-4}. \textit{Left:}
  $\xi >0$; \textit{right:} $\xi <0$. The color coding represents the Floquet exponent.
  The straight lines show an example of the time evolution of the coefficients for the
  zero and most excited $k_\textrm{max}$ modes at $\xi=30$.}
  \label{stab-chart}
\end{figure}

The slow variation of $\phi_0$ and the oscillation frequency can readily be taken into
account in conformal coordinates $\tau, Y_k$ defined by $dt=a\,d\tau $,
$Y_k\equiv a\,s_k$. Instead of~\eqref{Xk}, one has
\begin{equation}
  \ddot{ Y_k} + \left(   {k^2}  + \xi R a^2  - {\ddot a \over a} \right) \, Y_k=0 \;,
\label{Xk-tilde}
\end{equation}
where the dot now denotes differentiation with respect to $\tau$. In terms of the
conformal time, $R= 6 \ddot a /a^3$, which shows that $\xi=1/6$ implies no particle
production, as expected. At $\xi \gg 1$, the $\ddot a/a$ term can be neglected. The
inflaton amplitude decreases according to $\phi_0= \Phi_0/a$, where \(\Phi_0\) is the
initial inflaton field value. Then, following~\cite{Greene:1997fu}, one can define
$x\equiv\omega t = {1\over 2} \sqrt{\lambda_\phi} \Phi_0 \tau$ since
$\tau \propto \sqrt{t} \propto a$. In terms of the time variable $x$, one then recovers
Eq.~\eqref{ch-4} for $Y_k$ with $\kappa$ and $q$ rescaled by a factor of 4.

In the limit of constant coefficients, Eq.~\eqref{ch-4} belongs to the class of
ellipsoidal or Lam\'e wave equations. The solution can oscillate or grow in time
exponentially, depending on $\kappa $ and $ q$. The corresponding stability charts are
shown in Fig.~\ref{stab-chart}. The color coding represents the value of the Floquet
exponent $\mu$ such that the solution grows in time as $e^{\mu z}$ for a given set
$\kappa , q$. This resonant amplification of the amplitude is interpreted as particle
production.

For smaller momenta, particle production tends to be more efficient. However, the most
excited momentum is not zero. The coefficients of the ellipsoidal wave equation are not
exactly constant and evolve along straight lines in the $(q,\kappa)$ plane. As the system
evolves, they pass through areas with different Floquet exponents and the mode that gets
amplified the most over the course of this evolution dominates at the end of the
simulation. For moderate \(\xi\), we find
\begin{equation}
k_\textrm{max}^2 \simeq 0.8 \lambda_\phi \Phi_0^2 \;.
\end{equation}
The evolution of $\kappa,q$ for this and the zero modes is shown in Fig.~\ref{stab-chart}.

At weak coupling, the most excited momentum mode of the inflaton field can be read off
from the Lam\'e equation stability chart~\cite{Greene:1997fu}. In the absence of other
couplings, the system is conformal and the coefficients do not evolve in time, unlike in
the above example. The inflaton self-interaction excites the mode with
\begin{equation}
k_\textrm{max}^2 (\phi) \simeq 1.7 \lambda_\phi \Phi_0^2 \;.
\end{equation}
The occupation number of the excited momentum modes peaks initially at this value, while
subsequently the spectrum smoothens out due to rescattering.

We define the $k$-mode occupation numbers $n_k$ in the usual way. For example, in terms
of the $Y_k$ variables and conformal time $\tau$, $n_k $ is given by
\begin{align}
& n_k \equiv \frac{1}{2} \left( \omega_k |Y_k|^2 +\frac{1}{\omega_k} |\dot Y_k |^2 \right) \;, \\
& \omega^2_k \equiv k^2+a^2 \left\langle \frac{\partial^2 V(s)}{\partial s^2} \right\rangle \;.
\end{align}
As discussed above, this definition is meaningful in the Jordan frame only at late times,
when $\rho(s)$ in~\eqref{rho(s)} approaches the usual oscillator form. In this limit,
one can also drop the $\xi R$ term from the EOM of $Y_k$. Integrating $n_k$ over $k$,
one obtains the total particle number (see~\cite{Lebedev:2021tas} for precise definitions).

The DM self-interaction is included in the above considerations using the Hartree
approximation: it amounts to adding the mass-squared term proportional to
$\lambda_s \langle s^2 \rangle$ in the EOM for $s$. In the lattice formulation, however,
this approximation is not necessary and the full interaction term
${1\over 4} \lambda_s s^4$, which mixes the different momentum modes, is accounted for.

\subsection{End of resonance}\label{end}

In this subsection, we discuss the relevant time scales which determine the end of
resonant dark matter production. This depends strongly on the couplings and there are
several regimes. To understand the field dynamics in detail, we use classical lattice
simulations. These are reliable if the occupation numbers are large enough, which
corresponds to $\xi \gtrsim 5$ for positive $\xi$. We choose vacuum initial conditions
for the field fluctuations. These are mimicked by the Rayleigh probability distribution
for the momentum modes $|Y_k |$~\cite{Polarski:1995jg},
\begin{equation}
  P(Y_k) \propto \exp (-2 \omega_k | Y_k|^2) \;,
\end{equation}
where $\omega_k$ is the corresponding eigenfrequency at $t=0$. The phase of $Y_k$ is
assumed to have a random uniform distribution. For the inflaton field, we take the
initial conditions equivalent to
\begin{equation}
  \Phi_0 \simeq 0.9 \,M_\textrm{Pl}
\end{equation}
and zero initial velocity of $\phi$. In practice, it is more convenient to use a somewhat
lower initial value of $\phi$ and a larger $\dot \phi$ corresponding to the same energy
density. The simulations are performed with the numerical tool
CosmoLattice~\cite{Figueroa:2020rrl,Figueroa:2021yhd,cosm} customized to account for the
non-minimal coupling to gravity.\footnote{We are grateful to the authors
of~\cite{Figueroa:2021iwm} for providing this tool to us.} Related particle production
simulations in somewhat different regimes have been performed
in~\cite{Figueroa:2021iwm,Bettoni:2021zhq} and in~\cite{Ema:2017loe,Li:2022ugn} with
regard to the Higgs boson production.

In what follows, we define the end of the resonance as the period when the fast
exponential growth of the occupation numbers terminates. We focus on \(\xi>0\), although
many statements apply, at least qualitatively, to the \(\xi<0\) case as well modulo the
replacement \(\xi\to\vert\xi\vert\).

\subsubsection{Moderate \texorpdfstring{$\xi$}{ξ} and \texorpdfstring{$\lambda_s=0$}{λ s=0}}

In the range $5 \lesssim \xi \lesssim 30-40$, the resonance typically ends when the
$q$--parameter ($\propto a^{-2}$) becomes sufficiently small,
\begin{equation}
q \lesssim \mathcal{O}(1) \;,
\end{equation}
which suppresses the Floquet exponent to the level of $10^{-1}$ (Fig.~\ref{stab-chart}).
This occurs before backreaction of the produced quanta becomes strong enough to affect
the resonance. At moderate $\xi$, the resonance terminates when the scale factor is
$a \sim \mathcal{O}(\textrm{few})$, starting with $a_0=1$ at the beginning of the
simulation.

\begin{figure}[t!]
  \begin{center}
    \includegraphics[width=0.49\textwidth]{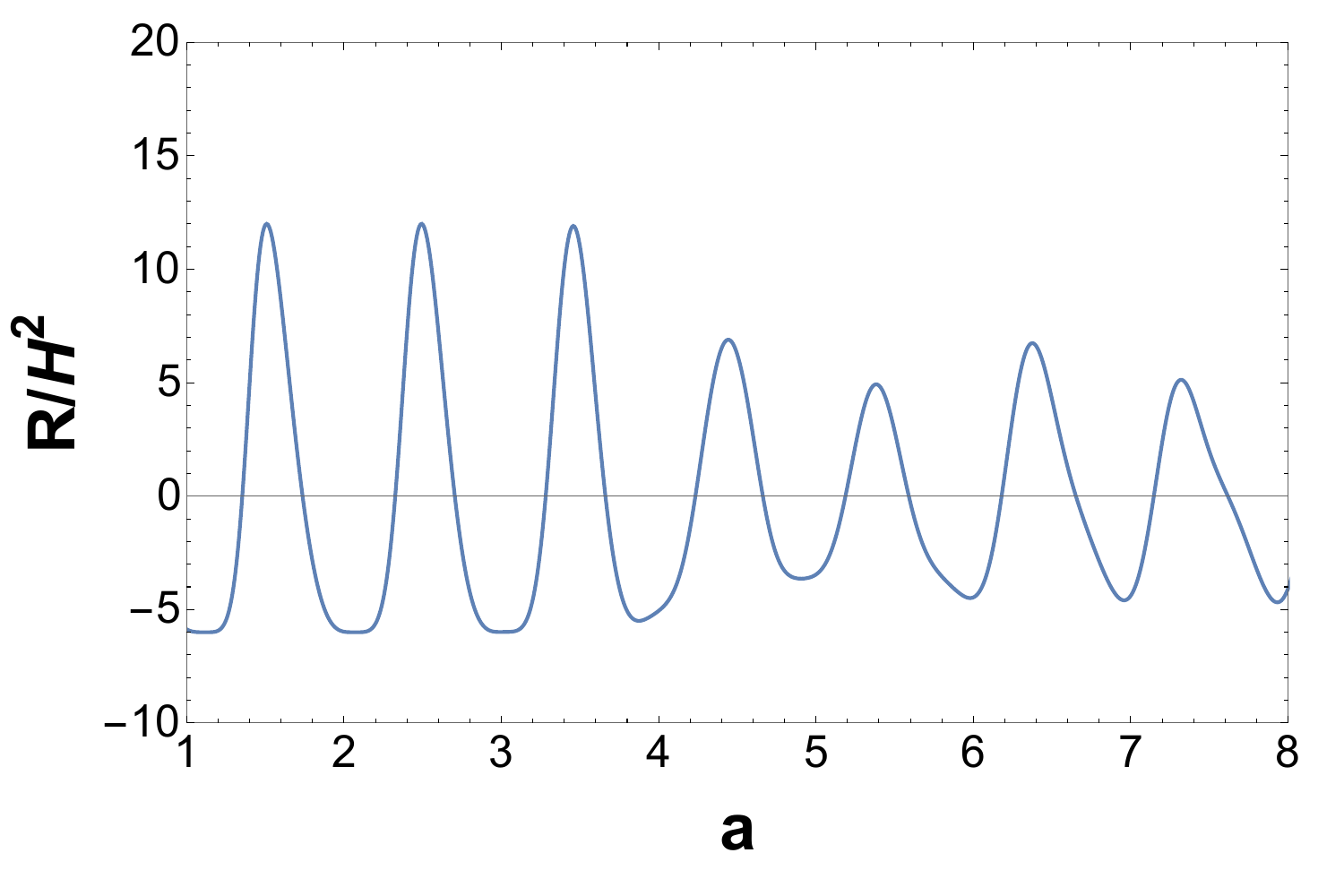}
    \includegraphics[width=0.49\textwidth]{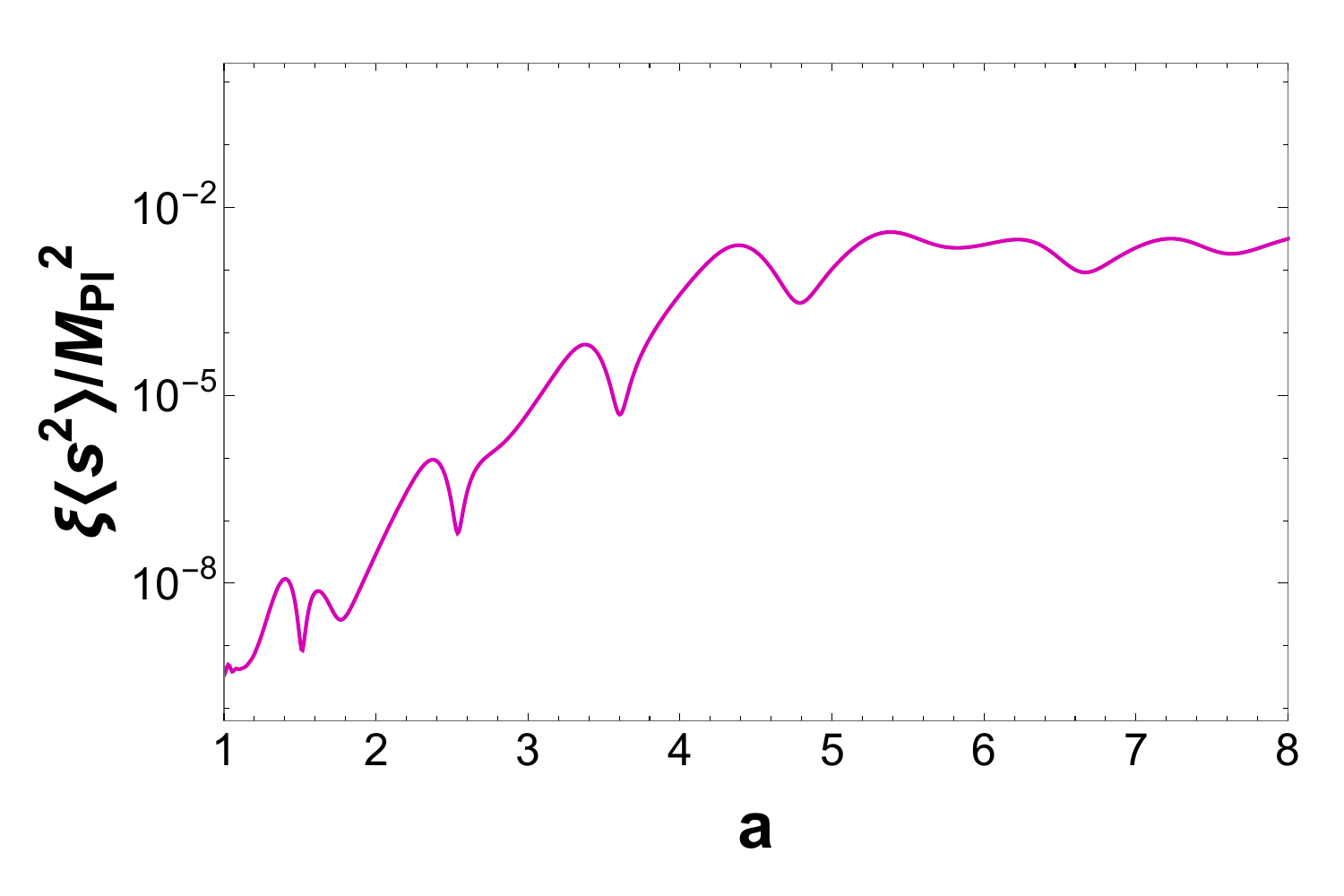}
  \end{center}
  \caption{Curvature and DM variance evolution in the quartic inflaton potential at
  $\xi=50$ and $\lambda_s=0$. The resonance terminates at $a\sim 4$.}
  \label{R-H-0}
\end{figure}

\subsubsection{Large \texorpdfstring{$\xi$}{ξ} and \texorpdfstring{$\lambda_s=0$}{λ s=0}}

For $\lambda_s=0$ and $\xi > 40$, the backreaction effects on the curvature become
essential and determine when the resonance terminates.

The evolution of the curvature is given by Eq.~\eqref{R-EOM}. Initially, it is fully
determined by the zero mode of the inflaton, $R \propto 4 V(\phi) - \dot \phi^2$, such
that $R$ exhibits regular oscillations. This leads to coherent DM production and
eventually the $\xi$--enhanced contribution of the latter to Eq.~\eqref{R-EOM} becomes
important. After that, the oscillations in $R$ become irregular (Fig.~\ref{R-H-0}) and
the resonance terminates due to backreaction of the produced particles on $R$. Thus, the
end of the resonance can be associated with
\begin{equation}
6 \xi \langle \partial^\mu s \partial_\mu s \rangle \sim 4 V(\phi) \;.
\end{equation}
Since the characteristic momentum squared for the tachyonic modes is $\xi R$
(see Eq.~\eqref{box-s}), while the right hand side is of order $M_\textrm{Pl}^2 R$, this
condition can be rewritten as
\begin{equation}
\sqrt{\langle s^2 \rangle}\sim {M_\textrm{Pl}\over \xi} \;.
\label{end-res}
\end{equation}
Note that, at this value of $\langle s^2 \rangle$, the function $F(s)$ starts to differ
from one. For very large $\xi \gtrsim 10^2$, the resonance is short-lived and terminates
before $a=2$. For instance, in the case $\xi=500$, the scalar curvature $R$ makes only
one oscillation before turning to zero, which signifies the onset of pure radiation
domination.

It is important to note that the above estimate does not give the terminal value of
$\langle s^2 \rangle$, which can increase further after the end of the resonance due to
strong rescattering effects. Numerically, we find that at $\xi \gtrsim 10^2$, the scalar
variance $\sqrt{\langle s^2 \rangle} $ tends to the value of order $10^{-2} M_\textrm{Pl}$
at $a\sim 10$.

\subsubsection{Large \texorpdfstring{$\xi$}{ξ} and significant \texorpdfstring{$\lambda_s$}{λ s}}

At small $\lambda_s$, the condition for the resonance termination remains as above,
while at stronger $\lambda_s$, the tachyonic resonance ends when the effective DM mass
turns positive. Since $s$-self-interaction induces effective mass squared of order
$\lambda_s \langle s^2 \rangle$ and the gravity-induced contribution is $\xi R$, we
obtain the following condition
\begin{equation}
\xi R \sim \lambda_s \langle s^2 \rangle \;.
\label{end-res-1}
\end{equation}
 The consequent $\langle s^2 \rangle$ is of order $\xi \lambda_\phi  \phi^4  / (\lambda_s M_\textrm{Pl}^2)$. This value is substantially smaller than that
in the $\lambda_s =0$ limit if
\begin{equation}
\lambda_s \gtrsim  \mathcal{O}( 10^{-2} \,\xi^3 \lambda_\phi) \;,
\end{equation}
where we have assumed that resonance terminates for $\phi$ a factor of a few below the
Planck scale.\footnote{The right hand side of the inequality contains the factor
$\phi^4 /M_\textrm{Pl}^4 $ and therefore sensitive to the exact value of~$\phi$.} Hence,
condition~\eqref{end-res-1} applies to this case.
\\ \ \\

\subsection{Dark matter relic abundance}\label{relic}

In this subsection, we study the relic abundance of dark matter generated via its
non-minimal coupling to gravity. To compute it, we need to model reheating, i.e.
conversion of the inflaton energy into the SM radiation. The simplest way to do it is to
introduce a small trilinear inflaton coupling to the Higgs field,
\begin{equation}
V_{\phi h} =  \sigma_{\phi h} \phi H^\dagger H \;.
\end{equation}
This leads to perturbative inflaton decay into the Higgs pairs, as long as
$m_\phi > 2 m_h$, and subsequent thermalization of the SM particle bath.\footnote{A pure
gravity-induced operator $\phi (H^\dagger H)^2 /M_{\rm Pl}$ also leads to the inflaton
decay  into the Higgses, however reheating via this operator is inefficient due to the
Planck suppression.} In this case, reheating occurs at late times, when the Jordan and
Einstein frames become essentially indistinguishable.

The dark matter abundance is characterized by
\begin{equation}\label{Y}
  Y = {n \over s_\textrm{SM} } ~~,~~ s_\textrm{SM} ={2\pi^2 \over 45} \, g_{*s} \, T^3 \;,
\end{equation}
where $n$ is the DM number density, $s_\textrm{SM}$ is the Standard Model entropy density
at temperature $T$ and $g_{*s}$ is the effective number of SM degrees of freedom
contributing to the entropy. In our case, DM interacts very weakly and never reaches
thermal equilibrium. Its particle number is approximately conserved after the preheating
stage and required to match the observed abundance~\cite{Planck:2015fie}
\begin{equation}
  Y_\infty =  4.4 \times 10^{-10} \; \left( {\textrm{GeV}\over m_s} \right) \;.
\end{equation}

Reheating takes place when
\begin{equation}\label{reh_cond}
  H_R \simeq \Gamma_{\phi\rightarrow hh} \;,\quad
  \Gamma_{\phi\rightarrow hh}=\frac{\sigma_{\phi h}^2}{8\pi m_\phi},
\end{equation}
where $H_R$ is the Hubble rate at reheating and $\Gamma_{\phi\rightarrow hh}$ includes
4 Higgs d.o.f.~at high energies. The reheating temperature is found via
\begin{equation}\label{HR}
  H_R=\sqrt{\frac{\pi^2g_*}{90}} \, {T_R^2 \over M_\textrm{Pl}} \;,
\end{equation}
where $g_* $ is the effective number of the SM degrees of freedom contributing to the
energy density. Combining the above $T_R$ and the dark matter number density $n$
computed with the lattice simulations, one determines $Y$ according to~\eqref{Y}.

We find that, in order to obtain the correct DM relic abundance, inflaton decay should
happen in the non-relativistic regime, i.e. when the average energy per inflaton quantum
is close to $m_\phi$. Therefore, the Universe evolution goes through a relativistic and
a non-relativistic expansion phase, prior to reheating,
\begin{equation}
  a_e\overset{\textrm{rel}}{\longrightarrow} a_*\overset{\textrm{nrel}}{\longrightarrow} a_R \;,
\end{equation}
where $a_e, a_*, a_R$ are the scale factors associated with the end of the simulation,
the onset of the non-relativistic regime and the reheating phase. In computing $a_R$, we
approximate the equation of state of the system by $w=1/3$ in the first period and $w=0$
in the second period. As a result, we obtain the following equation for $\sigma_{\phi h}$
producing the right DM relic density~\cite{Lebedev:2022ljz,Lebedev:2021tas}
\begin{equation}\label{sigma-DM-eq-1}
  \sigma_{\phi h} \simeq 5 \times 10^{-9} \; {m_\phi^{3/2} \over M_\textrm{Pl}^{1/2}} ~ {n_e(\phi) \over n_e(s)} ~~\left( {\textrm{GeV} \over m_s} \right) \;,
\end{equation}
which only requires the particle densities as an output of the simulations. Here
$n_e(\phi)$ and $n_e (s)$ are the particle number densities at the end of the simulation,
computed by integrating the occupation numbers for all the $k$-modes (including the zero
mode). At late times, the particle number and $n_e(\phi)/n_e (s)$ remain constant.

\begin{figure}[t!]
  \begin{center}
    \includegraphics[width=0.49\textwidth]{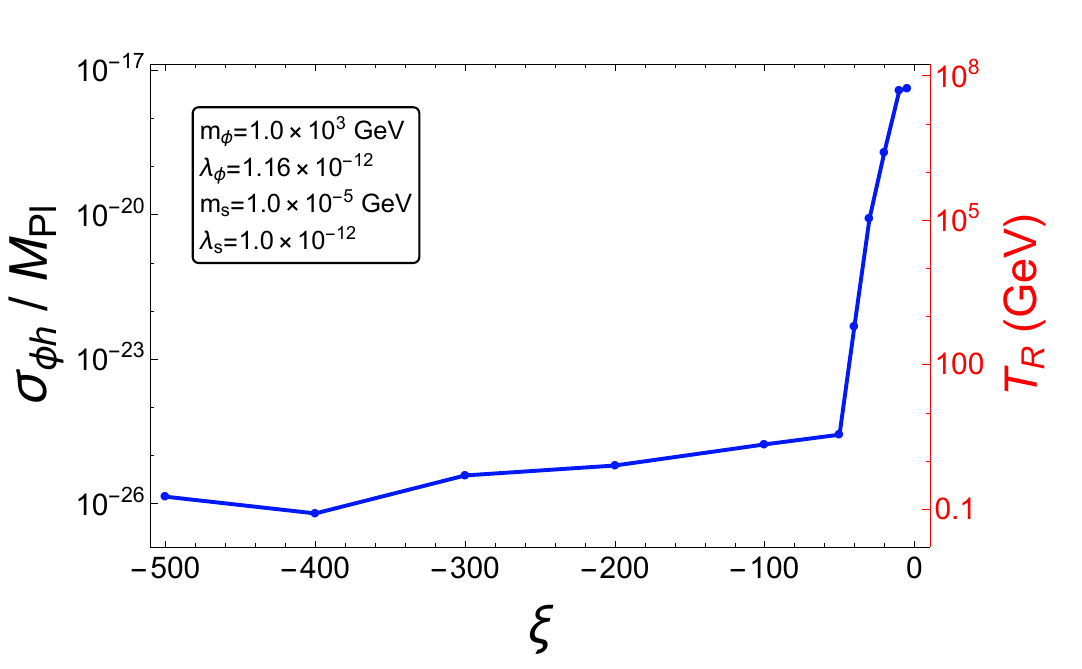}
    \includegraphics[width=0.49\textwidth]{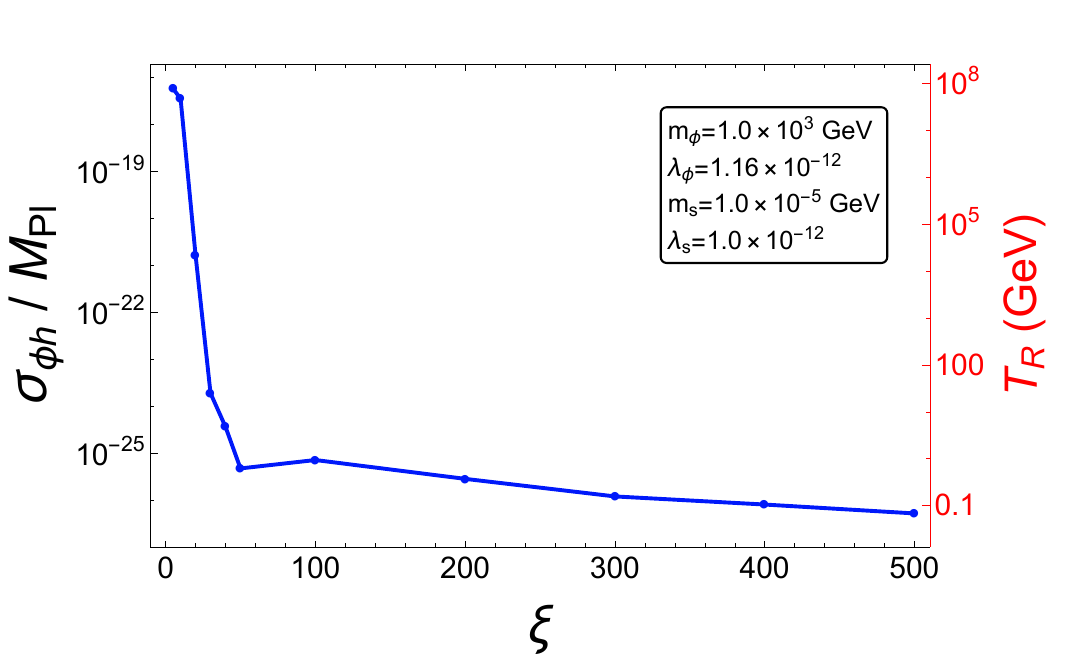}
  \end{center}
  \caption{The couplings producing the correct DM abundance in the quartic inflaton
  potential with the initial condition $\Phi_0 \simeq 0.9 \,M_\textrm{Pl}$. The area above
  the curve is ruled out by overabundance of dark matter. The reheating temperature can
  be read off from the $y$-axis on the right. The results for other DM masses are
  obtained by a simple rescaling according to~\eqref{sigma-DM-eq-1}.
  }
  \label{s-xi}
\end{figure}

Our numerical results for $\sigma_{\phi h}$ and $T_R$ are shown in Fig.~\ref{s-xi}. We
compute $n_e(\phi)$ and $n_e (s)$ with a version of CosmoLattice customised to include
the non-minimal coupling to gravity. The simulations yield the evolution of $a(t)$ and
$s(x), \phi(x)$ without resorting to the large $\xi$ approximation (as long as
$\xi \gtrsim 5$), which are used to calculate the $k$-mode occupation numbers at late
times. We find that these are well defined for $\lambda_s =10^{-12}$ and the DM particle
number is conserved in this regime.

For lower $\xi \sim 10$, the DM abundance is very sensitive to the exact value of the
non-minimal coupling. This is due to the exponential dependence of $n_e(s)$ on $\xi$.
However, such sensitivity is lost at higher $\xi$. We observe that the
$\sigma_{\phi h}(\xi)$ dependence flattens at $\xi>50$. This is characteristic of
quasi-equilibrium, i.e.~the state in which the average energy per inflaton and dark
matter quanta is roughly the same and $n_e(\phi)/n_e (s) \sim \mathcal{O}(1)$. This ratio
approaches one at around $\xi \sim 500$, while for yet higher values of $\xi$, the
simulations become unstable. Compared to the case of a direct inflaton-DM coupling
$\phi^2 s^2$~\cite{Lebedev:2021tas}, the system approaches exact quasi-equilibrium quite
slowly with respect to $\xi$. Indeed, the dynamics in the two cases are different: the
term $\xi s^2 R$ is active only for a short time, after which the distributions of
$\phi$ and $s$ evolve almost independently. In contrast, the direct coupling is active
for a longer period and equilibrates the two fields more efficiently. For completeness,
we also present our results for $\xi < 0$, which exhibit a similar trend to those for
$\xi >0$.

In Fig.~\ref{s-xi}, we choose a relatively low inflaton mass of 1 TeV. This is consistent
with the bound on the reheating temperature $T_R > 4\;$MeV~\cite{Hannestad:2004px} and
the DM relic abundance only for low enough $m_s$, which we take to be 10 keV. Particle
production after inflation is very intense such that larger $m_s$ would lead to DM
overabundance for our parameter choice. The required $\sigma_{\phi h}$ for different
$m_s$ can be obtained by a simple rescaling according to~\eqref{sigma-DM-eq-1}, as long
as $m_\phi^\textrm{} \gg m_s$. At large $\xi$ corresponding to the quasi-equilibrium
regime, the $m_\phi$-dependence also becomes simple: since $n_e(\phi)/n_e (s)\sim\mathcal{O}(1)$,
the coupling $\sigma_{\phi h}$ scales as $m_\phi^{3/2}$.

\begin{figure}[t!]
  \begin{center}
    \includegraphics[width=0.49\textwidth]{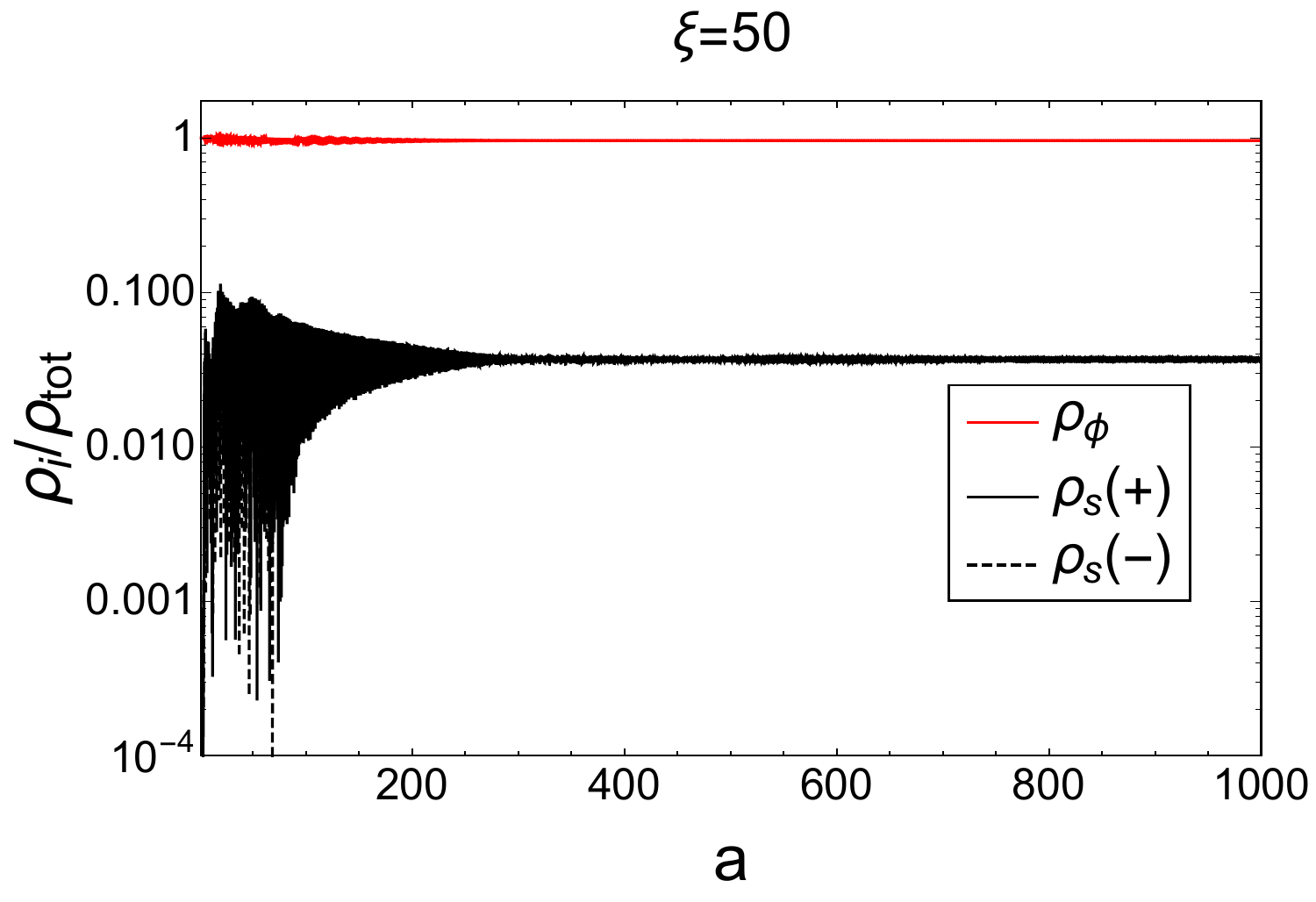}
    \includegraphics[width=0.49\textwidth]{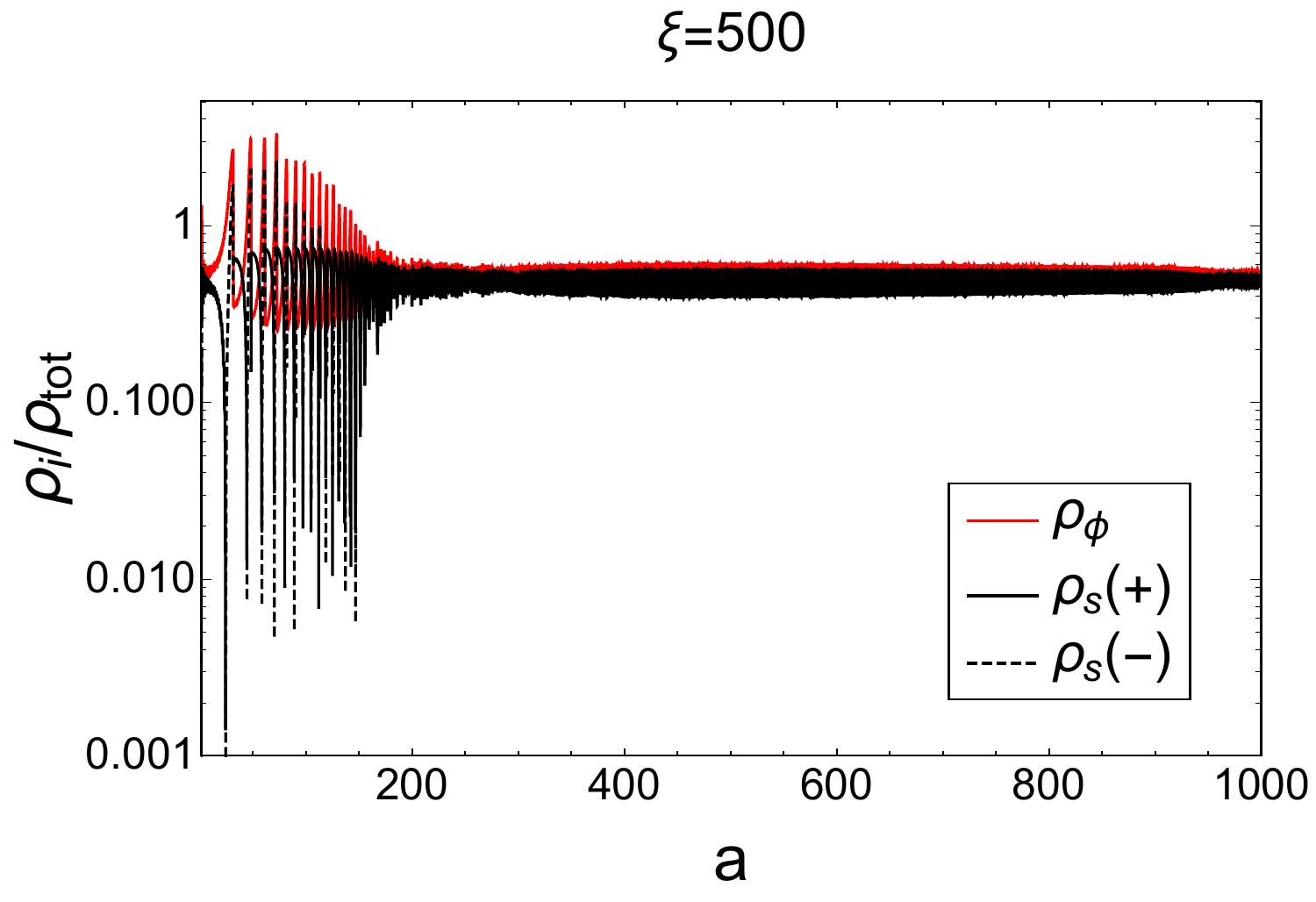}
  \end{center}
  \caption{Relative contributions of the inflaton ($\rho_\phi$) and dark matter ($\rho_s$)
  to the total energy density in the Jordan frame. At early times, $\rho_s$ can be negative
  (dashed line) due to the scalar-graviton mixing. At late times, the particle number is
  conserved.}
  \label{rho}
\end{figure}

Fig.~\ref{rho} illustrates the energy balance for $\xi=50$ and $\xi=500$. While DM
contributes about 4\% to the total energy at late times for $\xi=50$, this number goes up
to 50\% at $\xi=500$. We note that the DM energy density of $s$ can be negative at early
times due to the scalar-graviton mixing discussed in Section~\ref{jordan_dynamics}.

Finally, an important feature of our scenario is that the dark sector is much ``colder''
than the observable sector. This is due to the fact that the $s$-quanta are produced
immediately after inflation such that their momenta are subject to strong red-shifting.
We find, in particular, that the ratio between the SM bath temperature $T_\textrm{SM}$
and the characteristic energy of the DM quantum $ \langle E(s) \rangle $ after reheating
scales as
\begin{equation}
{T_\textrm{SM} \over \langle E(s) \rangle   } \sim \left(   {1\over \lambda_\phi} \, {a_R \over a_*}  \right)^{1/4} \gg 1 \;,
\end{equation}
where we have used the fact that the typical energy of the DM and inflaton quanta at
preheating is of order $\sqrt{\lambda_\phi} \phi_0$. Therefore, even very light DM
particles, well below a keV, become non-relativistic at the structure formation
temperature of order a keV. As a result, the structure formation constraints are easily
evaded.

\section{Quadratic local inflaton potential}

The inflaton potential around the minimum can also be dominated by the mass term,
\begin{equation}
V(\phi)\simeq {1\over 2} m_\phi^2 \, \phi^2\;.
\end{equation}
As $\phi(t)$ oscillates in the quadratic potential, dark matter quanta are produced
via tachyonic resonance. The analysis of particle production is analogous to that in the
$\phi^4$ case, although there are differences.

\subsection{The resonance}

In the $\phi^2$ potential, the inflaton field evolves according to
\begin{equation}
\phi(t) \simeq \phi_0 \, \cos m_\phi t \;,
\end{equation}
with slowly decreasing $\phi_0 \propto a^{-3/2}\propto (m_\phi t)^{-1}$. Thus, the
curvature has the form (see Eq.~\eqref{R-V})
\begin{equation}
R \simeq {\phi_0^2 m_\phi^2 \over 2 M_\textrm{Pl}^2} \,(3 \cos 2 m_\phi t +1) \;.
\end{equation}
At $\lambda_s=0$, Eq.~\eqref{Xk} then yields the EOM for the DM momentum modes $X_k$ in
the form of the Mathieu equation~\cite{Kofman:1997yn},
\begin{equation}
X_k^{\prime \prime} + \left[  A_k +2q \, \cos 4z\right]\, X_k=0 \;,
\end{equation}
where the $wH^2$ term has been neglected as before, the prime denotes differentiation
with respect to $z=m_\phi t/2$ , and
\begin{equation}
q= {3\xi \phi_0^2 \over M_\textrm{Pl}^2},~~A_k = {2\xi \phi_0^2 \over M_\textrm{Pl}^2} + {4k^2 \over a^2 m_\phi^2} \;.
\end{equation}
For the zero mode, one has $A_0/q = 2/3$. The large $q$ regime corresponds to broad
tachyonic resonance since the DM effective mass turns negative for the part of the
oscillation cycle. The DM self-interaction can be taken into account in this regime
using the Hartree approximation, as before, while lattice simulations allow us to go
beyond it.

\begin{figure}[t!]
  \begin{center}
    \includegraphics[width=0.49\textwidth]{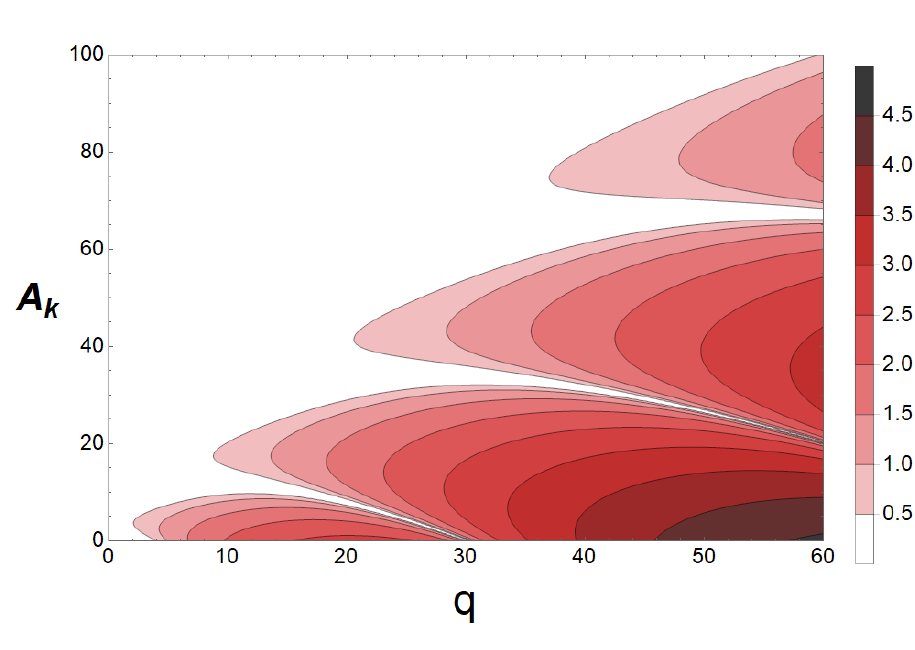}
    \includegraphics[width=0.49\textwidth]{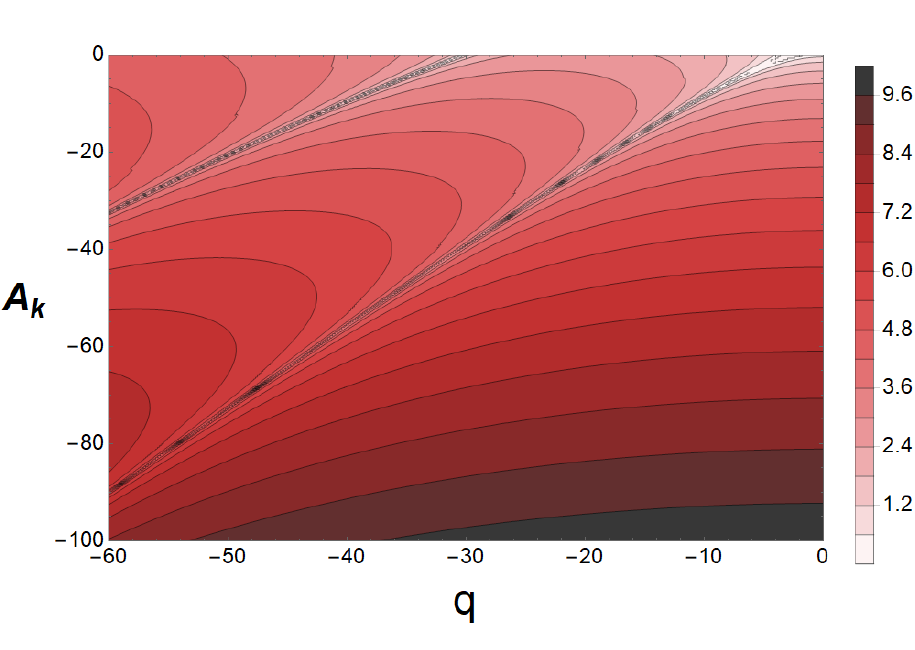}
  \end{center}
  \caption{Stability charts for the Mathieu equation. \textit{Left:} $\xi>0$;
  \textit{right:} $\xi<0$. The color coding represents the Floquet exponent.
  }
  \label{stab-chart-1}
\end{figure}

The stability chart for the Mathieu equation including the Floquet exponents is shown in
Fig.~\ref{stab-chart-1}. We observe that $q\sim 1$ leads to no exponential amplitude
growth. Unlike in the quartic case, the most excited modes are the infrared ones,
\begin{equation}
k_\textrm{max} \simeq 0 \;,
\end{equation}
which also trivially applies to the inflaton itself. We find, however, that the spectrum
of occupation numbers for tachyonic momentum modes is close to a flat one, with some IR
tilt.

\subsection{End of resonance}

To account for collective effects in the system dynamics, we perform lattice simulations
with CosmoLattice. We take the initial inflaton value to be
\begin{equation}
\Phi_0 \simeq 0.9 \, M_\textrm{Pl}
\end{equation}
and its initial velocity to be zero. The benchmark inflaton mass is set to
$0.6 \times 10^{-5}M_\textrm{Pl}$. We use $\lambda_s=10^{-6}$ to suppress the negative
energy contributions to $\rho_s$ at late times. Such a coupling does not lead to
$s$-thermalization for any DM mass above a keV~\cite{Arcadi:2019oxh}.

The considerations of Section~\ref{end} qualitatively apply to the quadratic case as well.
Focussing on the positive \(\xi\) case, we find that the simulations are reliable for
\(\xi>5\). The backreaction effects do not significantly affect the resonance for
\begin{equation}
\xi < 75 \;,
\end{equation}
in which case the resonance ends due to the time evolution of the Mathieu equation
coefficients, in particular, when $q$ becomes of order one. At larger $\xi$, the
resonance is shut off by backreaction on the curvature, as explained in Section~\ref{end},
which corresponds to
\begin{equation}
  \sqrt{\langle s^2 \rangle}  \sim {M_\textrm{Pl}\over \xi} \;.
\end{equation}
If, on the other hand, $\lambda_s $ is significant, the resonance can be terminated by
the induced DM mass when
\begin{equation}
  \lambda_s \langle s^2 \rangle \sim \xi R \;,
\end{equation}
which requires
\begin{equation}
  \lambda_s \gtrsim \mathcal{O} \left(   10^{-1} \,\xi^3 {m_\phi^2 \over M^2_\textrm{Pl}}  \right) \;.
\end{equation}
Here we have assumed that the resonance ends at $\phi$ a factor of a few below the Planck
scale. For our parameter choice, this inequality is translated to
$\lambda_s \gtrsim 10^{-6}$ at $\xi =100$. Since we set $\lambda_s = 10^{-6}$ as our
input value, the resonance is shut off by different mechanisms for $\xi$ above and below
100.

\subsection{Dark matter relic abundance}

\begin{figure}[t!]
  \begin{center}
    \includegraphics[width=0.49\textwidth]{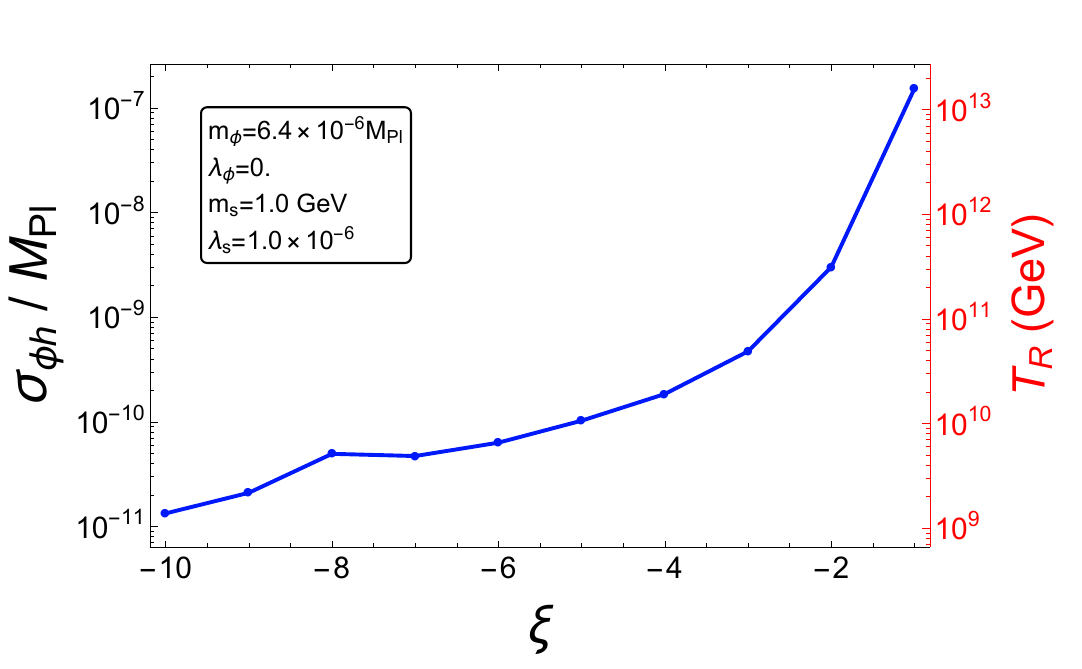}
    \includegraphics[width=0.49\textwidth]{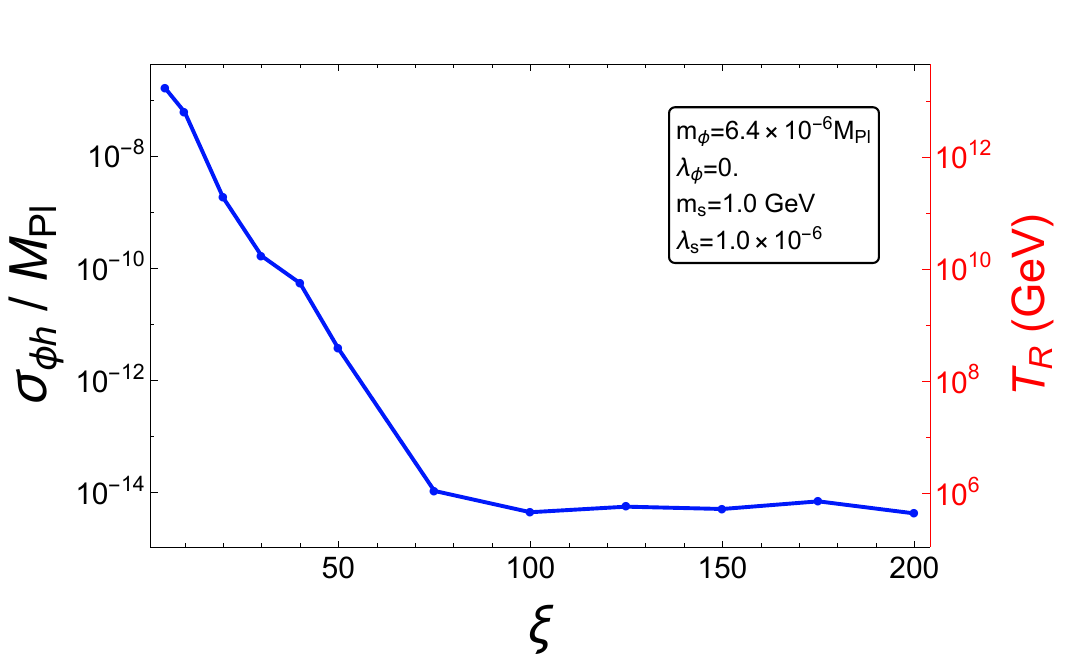}
  \end{center}
  \caption{Couplings and reheating temperature producing the correct DM abundance in the
  quadratic inflaton potential with $\Phi_0 \simeq 0.9 \,M_\textrm{Pl}$.~The area above
  the curve is ruled out by overabundance of dark matter. The results for other DM masses
  are obtained by a simple rescaling according to~\eqref{sigma-DM-eq-1}.}
  \label{s-xi-1}
\end{figure}

Repeating the steps put forth in Section~\ref{relic}, we compute the Higgs-inflaton
coupling $\sigma_{\phi h}$ and the reheating temperature required for the correct DM
relic abundance according to Eq.~\eqref{sigma-DM-eq-1}. The results are shown in
Fig.~\ref{s-xi-1}. The main difference from the quartic case is that the matter-dominated
expansion phase is much longer for the $\phi^2$ potential, which allows for higher
$\sigma_{\phi h}$ and $T_R$. For our parameter choice, the inflaton-DM system scales as
non-relativistic matter, so $a_e=a_*$.

For moderate $\xi$, the relic abundance is exponentially sensitive to the value of the
non-minimal coupling. At $\xi \sim 75$, the right panel of Fig.~\ref{s-xi-1} exhibits a
transition to qualitatively different behaviour. For larger $\xi$, the $\sigma_{\phi h}$
curve flattens and the system slowly approaches quasi-equilibrium, although
$n(\phi)/n(s) \sim 10^2$ still at this stage.

For completeness, in the left panel of Fig.~\ref{s-xi-1}, we present some of our results
for $\xi <0$. The simulation assumes vacuum initial conditions for the DM fluctuations,
which may not be a good approximation for negative $\xi$ in simple set-ups. Nevertheless,
this may be appropriate for more complicated multi-field inflationary models, where some
fields have direct couplings to $s$. We find that the classical dynamics approach is
valid for $\xi <-1$, yet we have not been able to obtain reliable results for $\xi < -10$
due to the large Floquet exponents and persistent presence of negative energies.

\begin{figure}[t!]
  \begin{center}
    \includegraphics[width=0.7\textwidth]{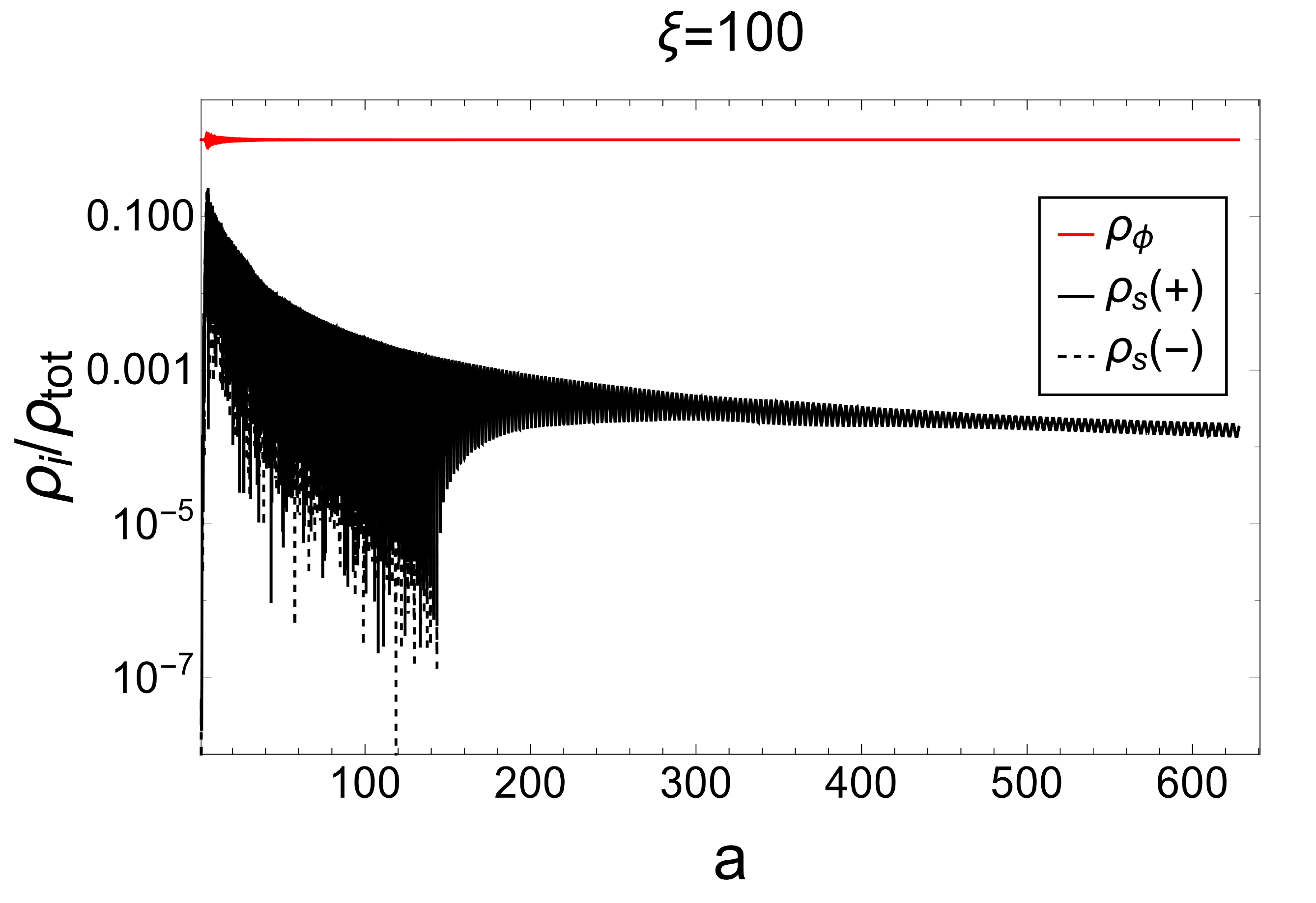}
  \end{center}
  \caption{Energy balance $\rho_i/\rho_\textrm{tot}$ in the inflaton-DM system at $\xi=100$,
  $\lambda_s=10^{-6}$. The dashed line represents negative $\rho_s$ appearing due to the
  scalar-graviton mixing. At late times, the particle number is conserved.}
  \label{rho-1}
\end{figure}

The negative energy issue becomes more significant in the $\phi^2$ potential case. This
is because the unwanted contribution to $\rho_s$ is proportional to $H\propto a^{-3/2}$,
which decreases slower than it did in the $\phi^4$ potential. As a result, a larger
$\lambda_s$ and/or longer running time is needed to obtain meaningful occupation numbers.
This tendency is seen in Fig.~\ref{rho-1}, which shows that negative energies appear
until $a\sim 150$ even for $\lambda_s=10^{-6}$. Eventually, the induced mass contribution
regularizes this behaviour since $\lambda_s \langle s^2 \rangle^2$ decreases slower than
$\xi H \langle s \dot s \rangle$ does. For $\xi<0$, on the other hand, there is an
additional negative contribution $\xi H^2\langle s^2\rangle$ to the DM energy density,
which makes the analysis more time consuming and computer-intensive.

We observe that, after the end of the resonance, dark matter takes up to 20\% of the
energy density of the system for $\xi=100$. Its contribution scales as radiation, so
this fraction decreases over time, unlike that in the $\phi^4$ case. As $\xi$ increases,
quasi-equilibrium is approached very slowly since the inter-species interaction is only
significant at early times. Although this trend is visible, we have not identified the
coupling $\xi$ necessary for the exact quasi-equilibrium, at least for the parameters at
hand.

It should be noted that our DM abundance predictions become less reliable at large
$\sigma_{\phi h}$, around $10^{-8} M_\textrm{Pl}$. At such coupling values, early time
Higgs production due to the tachyonic resonance becomes significant since
$\sigma_{\phi h} \Phi_0/m_{\phi}^2 \gg 1$. On the other hand, the energy transfer to the
Higgs field is small due to strong backreaction induced by the Higgs self-coupling
$\lambda_h \sim 10^{-2}$. It will further be diluted by matter-dominated expansion such
that we expect our results to give the right ballpark answer. Related lattice studies of
the $\sigma_{\phi h}$-induced resonance were performed in~\cite{Enqvist:2016mqj},
although with the assumption of $\lambda_h <0$ at high energy. This analysis would be
altered dramatically by backreaction for $\lambda_h \sim + 10^{-2}$, as an example
presented in~\cite{Lebedev:2021zdh} shows.

\section{Conclusion}

We have studied dark matter production via its non-minimal coupling to gravity $\xi s^2 R$
in both (locally) quadratic and quartic inflaton potentials. We find that collective
effects such as backreaction and rescattering make an important impact on the dark matter
abundance, especially at large couplings $\xi > 30$. The system tends to quasi-equilibrium
at yet larger $\xi > 100$, in which case the DM abundance becomes almost independent of
$\xi$. We also find that although the Jordan frame energy density of scalar DM can be
negative, its late time behaviour is sensible, at least in the presence of a small
self-coupling, which allows one to define a meaningful, conserved particle number in this
regime. The dark matter production mechanism is very efficient, originating from
tachyonic resonance, and leads to the correct DM abundance for a wide range of DM masses.

The mechanism considered in this paper constitutes one of possible gravitational channels
for dark matter production. The non-minimal coupling to curvature can be eliminated by a
metric transformation in favor of higher dimensional inflaton-dark matter couplings in
the Einstein frame. Such operators are also expected to be generated directly by quantum
gravitational effects~\cite{Lebedev:2022ljz,Lebedev:2022cic}, which creates a significant
uncertainty in the DM abundance calculations. In the present work, we focus entirely on
the effect of the non-minimal coupling to curvature, although a UV complete model would
have multiple sources of dark matter production.
\\ \ \\
\noindent
\textbf{Acknowledgements.}
We are grateful to the authors of~\cite{Figueroa:2021iwm}, especially D.~Figueroa and
A.~Florio, for providing us with an upgraded version of CosmoLattice. We also wish to
thank the Finnish Grid and Cloud Infrastructure (FGCI) for supporting this project with
computational and data storage resources. J.Y. would like to thank the lecturers and
participants of CosmoLattice School 2022 for sharing their expertise that assisted the
research presented in this work. J.Y. also acknowledges helpful discussions with Yann
Mambrini, Simon Cléry, and Essodjolo Kpatcha. We acknowledge support by Institut Pascal
at Université Paris-Saclay during the Paris-Saclay Astroparticle Symposium 2022, with the
support of the P2IO Laboratory of Excellence (program ``Investissements d'avenir''
ANR-11-IDEX-0003-01 Paris-Saclay and ANR-10-LABX-0038), the P2I axis of the Graduate
School of Physics of Université Paris-Saclay, as well as IJCLab, CEA, APPEC, IAS, OSUPS,
and the IN2P3 master projet UCMN.

\end{document}